# Conforming Discretizations of Boundary Element Solutions of the Electroencephalography Forward Problem

L. Rahmouni[a], S. B. Adrian[a,b], K. Cools[c], F. P. Andriulli[a]

[a]*Institut Mines-Télécom / Télécom Bretagne, Technopole Brest-Iroise, 29238, Brest, France*
[b]*Technische Universität München, Arcisstr. 21, 80333 Munich, Germany.*
[c]*The University of Nottingham, University Park, Nottingham, NG7 2RD, UK*

**Abstract**

In this paper we present a new discretization strategy for the boundary element formulation of the Electroencephalography (EEG) forward problem. Boundary integral formulations, classically solved with the Boundary Element Method (BEM), are widely used in high resolution EEG imaging because of their recognized advantages, in several real case scenarios, in terms of numerical stability and effectiveness when compared with other differential equation based techniques. Unfortunately however, it is widely reported in literature that the accuracy of standard BEM schemes for the forward EEG problem is often limited, especially when the current source density is dipolar and its location approaches one of the brain boundary surfaces. This is a particularly limiting problem given that during an high-resolution EEG imaging procedure, several EEG forward problem solutions are required for which the source currents are near or on top of a boundary surface.

This work will first present an analysis of standardly and classically discretized EEG forward problem operators, reporting on a theoretical issue of some of the formulations that have been used so far in the community. We report on the fact that several standardly used discretizations of these formulations are consistent only with an $L^2$-framework, requiring the expansion term to be a square integrable function (i.e., in a Petrov-Galerkin scheme with expansion and testing functions). Instead, those techniques are not consistent when a more appropriate mapping in terms of fractional order Sobolev spaces is considered. Such a mapping allows the expansion function term to be a less regular function, thus sensibly reducing the need for mesh refinements and low-precisions handling strategies that are currently required. These more favorable mappings, however, require a different and conforming discretization which must be suitably adapted to them. In order to appropriately fulfill this requirement, we adopt a mixed discretization based on dual boundary elements residing on a suitably defined dual mesh. We devote also a particular attention to implementation-oriented details of our new technique that will allow the rapid incorporation of our finding in one's own EEG forward solution technology. We conclude by showing how the resulting forward EEG problems show favorable properties with respect to previously proposed schemes and we show their applicability to real case modeling scenarios obtained from Magnetic Resonance Imaging (MRI) data.

State of the art high resolution Electroencephalography (EEG) can righteously be considered a fully fledged imaging technique for the brain (Acar et al., 2016). Its high temporal resolution, together with the compatibility and complementarity with other imaging strategies (Magnetoencephalography (MEG), Positron Emission Tomography (PET), and Magnetic Resonance Imaging (MRI)) (Siems et al., 2016; Peng et al., 2016; Dabek et al., 2015; Bénar et al., 2007), explains the steady interest that EEG is attracting in neuroimaging (Huang et al., 2015; Jorge et al., 2015; Fiederer et al., 2016). The peculiarity of high resolution EEGs, with respect to the traditional analyses based on grapho-elements, is the reconstruction

*Email addresses:* lyes.rahmouni@telecom-bretagne.eu (L. Rahmouni), simon.adrian@tum.de (S. B. Adrian), Kristof.Cools@nottingham.ac.uk (K. Cools), francesco.andriulli@mines-telecom.fr (F. P. Andriulli)

of the volume brain sources based on scalp potential data (Phillips et al., 2002; Koessler et al., 2010). This is the EEG inverse source problem, which is, as it is well-known, ill-posed (Grech et al., 2008). The solution of the EEG inverse source problem relies on multiple iterated solutions of the EEG forward problem where, known the configuration of brain sources, the electric potential is recovered at the scalp (Pascual-Marqui, 1999). The accuracy in the solution of the EEG forward problem clearly impacts and limits the accuracy of the associated EEG inverse problem: a low accuracy of the solutions of the EEG forward problem translates in a low accuracy of the inverse problem solution (Acar and Makeig, 2013). This results in the pressing need to keep the accuracy of the EEG forward problem as high as possible.

Among the techniques to solve the EEG forward problem, Boundary Element Method (BEM) is a widely used one (Hallez et al., 2007a). This numerical strategy is based on an integral formulation equivalent to the Poisson equation and, when compared with other numerical approaches like the Finite Element Method (FEM) or the Finite Difference Method (FDM) (Hallez et al., 2007b), BEM based solvers only discretize the surfaces enclosing the different brain regions and do not require the use of boundary conditions to terminate the solution domain. This results in interaction matrices of a smaller dimensionality (He et al., 1987) and explains the popularity of the BEM approach in the scientific community. Unfortunately, standard BEM methods are no panacea. It is widely reported, in fact, that the accuracy of standard BEM schemes for the forward EEG problem is often limited, especially when the current source density is dipolar and its location approaches one of the brain boundary surfaces (Fuchs et al., 2001; He et al., 1999). This is a particularly limiting problem given that during the solution of the EEG inverse source problem, several forward EEG problem solutions are required for which the the primary current density terms are near or on top of a boundary surface (Cosandier-Rimélé et al., 2008; Fuchs et al., 1998).

Three main strategies have been reported in literature to limit the impact of accuracy losses: (i) the avoidance of brain source modeling near boundaries (Yvert et al., 2001), (ii) the use of global or local mesh refinements that can better handle the singularity of the dipolar source term (Meijs et al., 1989; Zanow and Peters, 1995; Fuchs et al., 1998), and (iii) the introduction of a symmetric boundary element formulation (Adde et al., 2003; Kybic et al., 2005). All the above mentioned techniques can sensibly improve source-related precision issues, but at the same time they present some undesirable drawbacks: (i) avoiding the positioning of dipolar sources near boundaries on one hand represents a limitation on correct modeling (Cosandier-Rimélé et al., 2008) and on the other hand it increases the ill-posedness of the inverse-source problem (Scherzer, 2011). (ii) The use of mesh refinements increases the computational burden, due to the higher dimensionality of the refined models and this can result in substantial inefficiencies (Yvert et al., 2001; Wolters et al., 2007). This is especially true in the context of inverse source problem solutions, where sources are often equally distributed near the boundaries of brain layers (Cosandier-Rimélé et al., 2008). (iii) The use of symmetric formulations, that are based on a clever and complete exploitation of the representation theorem, results in the simultaneous solution of two integral equations in two unknowns and sensibly improves the accuracy of BEM method based EEG imaging. However, these formulations results in more unknowns, which increases the computational complexity of the EEG forward and inverse solutions. Moreover, the symmetric formulation in (Adde et al., 2003; Kybic et al., 2005) presents a conditioning that is dependent and growing with the number of unknowns (or equivalently with the inverse of the mesh parameter). This ill-conditioning results in harder-to-obtain numerical solutions for realistic problems as the matrix inversion becomes an increasingly unstable operation (Sauter and Schwab, 2011).

To circumvent the above mentioned limitations, this work proposes a different approach. We first start from analyzing the mapping properties of standard EEG forward problem operators (double and adjoint double layer). We report on the fact that standardly used discretizations of these operators are consistent only with an $L^2$-formulation, requiring the expansion term to be a square integrable function. Instead, those techniques are not consistent when a mapping in terms of fractional order Sobolev spaces is considered. Such a mapping, in the case of the adjoint double layer operator, would allow the expansion term to be a less regular function, sen-sibly reducing the need for mesh refinements and low-precisions handling strategies currently required. These more favorable mappings, however, require a different and conforming discretization which must be suitably adapted to them. Some of the authors of this work presented in the past a strategy to comply with proper Sobolev space mappings based on dual elements. This approach was introduced in (Cools et al., 2009) and named "mixed discretization". Mixed discretizations are conforming with respect to Sobolev properties of second kind operators. This approach has been subsequently applied to several problems in electromagnetics (Cools et al., 2011; Yan et al., 2011) and acoustics (Ylä-Oijala et al., 2015). In this work we have applied the mixed discretization concept to the case of multi-layered EEG operators used to solve piecewise homogeneous and isotropic nested head models. This discretization strategy can be extended to non nested topologies. The resulting forward EEG problems show favorable properties with respect to previously proposed schemes. As complement to the theoretical and numerical treatments, a particular attention has been devoted to implementation-oriented details that will allow the specialized practitioner to easily incorporate these findings in his EEG forward solution technology. Very preliminary and partial results of this contribution have been presented in a conference contribution (Rahmouni and Andriulli, 2014).

This paper is organized as follows: in Section 1 we first review classical EEG discretizations and we analyze their consistency with respect to fractional order Sobolev space mappings; we then introduce dual basis functions and the new forward EEG mixed discretized formulations we propose in this work. In Section 2, we present a complete numerical study of the new techniques to comparatively test their performance against the state of the art. This will be done on both canonical spherical models (for which benchmarking against analytic solutions is possible) and on realistic models arising from MRI data. Section 3 presents our discussion of these results and our conclusions.

## 1. Methods

### 1.1. Standard Integral Equation Formulations of the Electroencephalography Forward Problem

Let $\sigma$ be a smooth, isotropic conductivity distribution and let $\boldsymbol{j}$ be a quasi-static electric volume current density distribution in $\mathbb{R}^3$. The current density $\boldsymbol{j}$ generates the electric potential $\phi$, a relationship that is mathematically expressed by the Poisson's equation

$$\nabla \cdot \sigma \nabla \phi = f = \nabla \cdot \boldsymbol{j}, \quad \text{in } \mathbb{R}^3. \tag{1}$$

When $\sigma$ models the conductivity distribution of a human head, the problem of finding the electric potential $\phi$ is denoted as the EEG forward problem (Hallez et al., 2007a; Grech et al., 2008).

In BEM techniques, the head is usually modeled by domains of different areas of constant conductivity. The conductivity $\sigma$ is a piecewise constant function dividing the space $\mathbb{R}^3$ in a nested sequence of regions as depicted in Fig. 1. The different domains corresponding to the regions where $\sigma$ is constant and equal to $\sigma_i$ are labeled $\Omega_i$ with $i = 1, \ldots, N + 1$. The domain $\Omega_{N+1}$ is the exterior region, extending to infinity, with $\sigma_{N+1} = 0$. In $\Omega_{N+1}$ no current sources are present. The surfaces separating the different regions of conductivity are labeled $\Gamma_i$ with $i = 1, \ldots, N$ as shown in Fig. 1.

In order to account for piecewise continuous $\sigma$, Eq. (1) must be complemented by transmission and boundary conditions resulting in (Pruis et al., 1993)

$$\sigma_i \Delta \phi = f \quad \text{in } \Omega_i, \quad \text{for all } i = 1, \ldots, N, \tag{2}$$

$$\Delta \phi = 0 \quad \text{in } \Omega_{N+1}, \tag{3}$$

$$[\phi]_j = 0 \quad \text{on } \Gamma_j, \text{ for all } j = 1, \ldots, N, \tag{4}$$

$$[\sigma \partial_{\hat{\boldsymbol{n}}} \phi]_j = 0 \quad \text{on } \Gamma_j, \text{ for all } j = 1, \ldots, N. \tag{5}$$

The expression $[g]_j$ denotes the jump of the function $g$ at the surface $\Gamma_j$, that is,

$$[g]_j = g|_{\Gamma_j}^- - g|_{\Gamma_j}^+, \tag{6}$$

with $g|_{\Gamma_j}^-$ and $g|_{\Gamma_j}^+$ the interior and exterior limits of $g$ at the surface $\Gamma_j$, respectively. These limits are defined as

$$g|_{\Gamma_j}^\pm (\boldsymbol{r}) := \lim_{\alpha \to 0^\pm} g(\boldsymbol{r} + \alpha \hat{\boldsymbol{n}}) \quad \text{for all } \boldsymbol{r} \text{ on } \Gamma_j, \tag{7}$$

where $\hat{\boldsymbol{n}}$ denotes the normal at each surface (see Fig. 1).

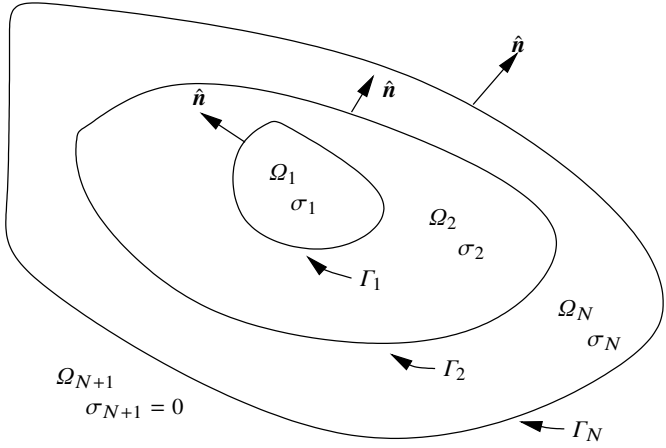

Figure 1: Nested sequences of regions with constant conductivity.

*1.1.1. Boundary Integral Operators*

Boundary element methods provide a numerical approximation of the potential $\phi$ (Huiskamp et al., 1999; Steinbach, 2008) when the forward EEG problem is cast in an integral equation formulation. In the following, we introduce the integral operators and their mapping properties, and we review the standard integral formulations of the EEG forward problem.

**Definition 1** (Boundary Integral Operators). Let $\Omega \subset \mathbb{R}^3$ be a bounded Lipschitz domain with boundary $\Gamma := \partial\Omega$. We define the single layer operator

$$\mathcal{S} : H^{-1/2}(\Gamma) \to H^{1/2}(\Gamma),$$
$$(\mathcal{S}u)(\boldsymbol{r}) = \int_\Gamma G(\boldsymbol{r} - \boldsymbol{r}')u(\boldsymbol{r}')\mathrm{d}S(\boldsymbol{r}'), \quad (8)$$

the double layer and adjoint double layer operator

$$\mathcal{D} : H^{1/2}(\Gamma) \to H^{1/2}(\Gamma),$$
$$(\mathcal{D}u)(\boldsymbol{r}) := \int_\Gamma \partial_{\hat{\boldsymbol{n}}'} G(\boldsymbol{r} - \boldsymbol{r}')u(\boldsymbol{r}')\mathrm{d}S(\boldsymbol{r}'), \quad (9)$$

$$\mathcal{D}^* : H^{-1/2}(\Gamma) \to H^{-1/2}(\Gamma),$$
$$(\mathcal{D}^*u)(\boldsymbol{r}) = \int_\Gamma \partial_{\hat{\boldsymbol{n}}} G(\boldsymbol{r} - \boldsymbol{r}')u(\boldsymbol{r}')\mathrm{d}S(\boldsymbol{r}'), \quad (10)$$

and the hypersingular operator

$$\mathcal{N} : H^{1/2}(\Gamma) \to H^{-1/2}(\Gamma),$$
$$(\mathcal{N}u)(\boldsymbol{r}) = \int_\Gamma \partial_{\hat{\boldsymbol{n}},\hat{\boldsymbol{n}}'} G(\boldsymbol{r} - \boldsymbol{r}')u(\boldsymbol{r})\mathrm{d}S(\boldsymbol{r}'). \quad (11)$$

In the definitions above, the function

$$G(\boldsymbol{r} - \boldsymbol{r}') = \frac{1}{4\pi|\boldsymbol{r} - \boldsymbol{r}'|} \quad (12)$$

is the free-space Green's function. The Sobolev spaces $H^x$, $x \in \{-1/2, 1/2\}$, appearing in the mapping properties are briefly defined in Appendix A.

*Remark.* The reader should be warned that there is no consistent naming of the operators above in the literature and the naming choice made here is the one classically adopted in potential theory (see for example (Sauter and Schwab, 2011)).

*Source Modeling and Inhomogeneous Solution.* Current dipoles are a common approximation of brain electric sources making them a widely used model in the forward and inverse EEG problem (De Munck et al., 1988; Sarvas, 1987; Schimpf et al., 2002). The current dipole is defined by

$$\boldsymbol{j}_{\mathrm{dip}}(\boldsymbol{r}) = \boldsymbol{q}\delta_{\boldsymbol{r}_0}(\boldsymbol{r}) \quad (13)$$

where $\boldsymbol{q}$ represents the dipole moment and $\delta_{\boldsymbol{r}_0}$ the Dirac delta function. The corresponding potential in an infinite homogeneous domain is

$$v_{\mathrm{dip}}(\boldsymbol{r}) = \frac{1}{4\pi}\frac{\boldsymbol{q} \cdot (\boldsymbol{r} - \boldsymbol{r}_0)}{|\boldsymbol{r} - \boldsymbol{r}_0|^3}. \quad (14)$$

Throughout the following sections, we use

$$v_{\mathrm{s},\Omega_i} = v_{\mathrm{dip}} \quad \text{for } \boldsymbol{r}_0 \in \Omega_i. \quad (15)$$

Moreover, whenever two underscore indices $j, i$ are added to an operator symbol we mean that, in defining the operator, the integration is constrained to the $i$th surface and the integral is evaluated only on the $j$th surface. For example, $\mathcal{S}_{ji}$ is defined as

$$(\mathcal{S}_{ji}p)(\boldsymbol{r}) = \int_{\Gamma_i} G(\boldsymbol{r} - \boldsymbol{r}')p(\boldsymbol{r}')\mathrm{d}S(\boldsymbol{r}'), \quad \boldsymbol{r} \in \Gamma_j. \quad (16)$$

*1.1.2. Boundary Integral Formulations*

Three integral formulations are commonly used for computing the electric potential $\phi$ in Eq. (1) (Kybic et al., 2005; Stenroos and Sarvas, 2012; Gramfort et al., 2014; Birot et al., 2014). All of them leverage the same principle: the electric potential $\phi$ is decomposed into

$$\phi = v_\mathrm{s} + v_\mathrm{h} \quad (17)$$

such that $\sigma_i \Delta v_s = f$ in $\Omega_i$ for all $i = 1, \ldots, N$ (see Eq. (2)) and such that $v_h$ is a piecewise harmonic correction ensuring that $\phi$ will satisfy the boundary conditions (4) and (5). For setting the notation and for the sake of self-consistency, we list these formulations below; for a more detailed derivation, we refer the reader to (Kybic et al., 2005) and references therein.

*The adjoint double layer formulation.* In this formulation, the ansatz for $v_{s1}$ has the following form:

$$v_{s1} = \sum_{i=1}^{N} \frac{v_{s,\Omega_i}}{\sigma_i}. \quad (18)$$

This choice satisfies Eq. (2) and Eq. (3), and in addition, $[v_{s1}]_j = 0$ and $[\partial_{\hat{n}} v_{s1}]_j = 0$. Theorem 1 in Appendix B is then used to construct a harmonic function a $v_{h1}$ such that a Neumann's boundary condition is satisfied. It is obtained that

$$\partial_{\hat{n}} v_{s1}|_{\Gamma_j} = \frac{\sigma_j + \sigma_{j+1}}{2(\sigma_{j+1} - \sigma_j)} q_{\Gamma_i} - \sum_{i=1}^{N} \mathcal{D}^*_{ji} q_{\Gamma_i} \quad \text{for } i = 1, \ldots, N. \quad (19)$$

*The double layer formulation.* The following particular solution is put forward (see, for example, Kybic et al. (2005) or Stenroos and Sarvas (2012))

$$v_{s2} = \sum_{i=1}^{N} v_{s,\Omega_i}, \quad (20)$$

which satisfies Eq. (2), $[v_{s2}] = 0$, and $[\partial_{\hat{n}} v_{s2}] = 0$. After complementing it with a harmonic solution $v_{h2}$ that satisfy $[\partial_{\hat{n}} v_{h2}] = 0$, it is obtained

$$v_{s2}|_{\Gamma_j} = \frac{\sigma_j + \sigma_{j+1}}{2} \phi_{\Gamma_i} - \sum_{i=1}^{N} (\sigma_{i+1} - \sigma_i) \mathcal{D}_{ji} \phi_{\Gamma_i}. \quad (21)$$

*The symmetric Formulation.* Differently from the previous two approaches, in the symmetric formulation, the harmonic function $v_{h3}$ is constructed as follows (Kybic et al., 2005)

$$v_{h3,\Omega_i} = \begin{cases} \phi - \frac{v_{s,\Omega_i}}{\sigma_i}, & \text{in } \Omega_i, \\ -\frac{v_{s,\Omega_i}}{\sigma_i}, & \text{in } \mathbb{R}^3 \setminus \overline{\Omega_i}. \end{cases} \quad (22)$$

Then, it can be shown that

$$\sigma_{i+1}^{-1}(v_{s,\Omega_{i+1}})|_{\Gamma_i} - \sigma_i^{-1}(v_{s,\Omega_i})|_{\Gamma_i}$$
$$= \mathcal{D}_{i,i-1}\phi_{\Gamma_{i-1}} - 2\mathcal{D}_{ii}\phi_{\Gamma_i} + \mathcal{D}_{i,i+1}\phi_{\Gamma_{i+1}}$$
$$- \sigma_i^{-1}\mathcal{S}_{i,i-1}d_{\Gamma_{i-1}} + (\sigma_i^{-1} + \sigma_{i+1}^{-1})\mathcal{S}_{ii}d_{\Gamma_i} - \sigma_{i+1}^{-1}\mathcal{S}_{i,i+1}d_{\Gamma_{i+1}}$$
$$(23)$$

and

$$(\partial_{\hat{n}} v_{s,\Omega_{i+1}})|_{\Gamma_i} - (\partial_{\hat{n}} v_{s,\Omega_i})|_{\Gamma_i}$$
$$= \sigma_i \mathcal{N}_{i,i-1}\phi_{\Gamma_{i-1}} - (\sigma_i + \sigma_{i+1})\mathcal{N}_{ii}\phi_{\Gamma_i} + \sigma_{i+1}\mathcal{N}_{i,i+1}\phi_{\Gamma_{i+1}}$$
$$- \mathcal{D}^*_{i,i-1}d_{\Gamma_{i-1}} + 2\mathcal{D}^*_{ii}d_{\Gamma_i} - \mathcal{D}^*_{i,i+1}d_{\Gamma_{i+1}} \quad (24)$$

hold for $i = 1, \ldots, N$. Here we have used the notation $d_{\Gamma_i} = \sigma_i \partial_{\hat{n}} \phi|_{\Gamma_i}^-$

*Isolated Skull Approach.* In the presence of a layer of low conductivity, the double layer formulation suffers from numerical inaccuracies. To overcome this problem, Hamalainen and Sarvas (1989) proposed a numerical strategy named Isolated Skull Approach. This scheme was first formulated for a three layered head model and then generalized to an arbitrary number of layers by Meijs et al. (1989); Gençer and Akalin-Acar (2005). The principle of this approach is to write the total potential as a sum of two terms:

$$\phi = \phi_{\text{ISA}} + \phi_{\text{corr}}. \quad (25)$$

The first term $\phi_{\text{ISA}}$ is the potential computed assuming an isolated model consisting of only the compartments which are under the skull. The second part $\phi_{\text{corr}}$ is a correction term computed as in the standard double layer formulation with a right hand side equal, for a three layers model, to

$$v_{s3} = \sigma_{\text{skull}}(v_{s2} + \phi_{\text{ISA}}), \quad (26)$$

where $\sigma_{\text{skull}}$ is the conductivity of the skull. We note that $\phi_{\text{ISA}}$ is different from zero only on surfaces corresponding to tissues located under the skull.

The isolated skull approach and the double layer formulation use the same operator, namely $\mathcal{D}$; hence, both schemes have the same requirements in term of mapping properties.

*1.2. Analysis of the Main Drawbacks of Standard Discretizations*

To evidence the drawbacks of standard (currently used in literature) BEM discretizations of the EEG forward problem, we have to consider Petrov-Galerkin theory, which provides the convergence properties of a numerical boundary element solution in the case of asymmetric discretizations (Steinbach, 2008).

### 1.2.1. Petrov-Galerkin Method Reviewed

Let $X$ and $Y$ be Hilbert spaces, and $\mathcal{A} : X \to Y'$ a bounded, linear operator. We can associate with $\mathcal{A}$ a bilinear form $a : X \times Y \to \mathbb{R}$ that satisfies $|a(x, y)| \leq C \|x\|_X \|y\|_Y$ with $C > 0$. We are faced with the variational problem to find $u \in X$ such that $a(u, v) = \langle f, v \rangle_{Y' \times Y}$ for all $v \in Y$ with $f \in Y'$.

To solve this variational formulation, we cast this problem into a matrix-vector equation by using finite-dimensional subspaces $X_h \subset X$ and $Y_h \subset Y$ with $\dim(X_h) = \dim(Y_h) = M$. The task is to find $u_h \in X_h$ such that $a(u_h, v_h) = \langle f, v_h \rangle_{Y'_h \times Y_h}$ for all $v_h \in Y_h$. The function $u_h$ is an approximation of $u$ and this approach is called Petrov-Galerkin method (Steinbach, 2008).

The key ingredient of the Petrov-Galerkin method is that the testing is always performed in the dual space of the range of $\mathcal{A}$, where we notice that $Y'' = Y$ because of the reflexivity of $Y$.

### 1.2.2. Standard Discretizations

The classical integral equations presented in Section 1.1.2 are discretized using a BEM approach. The different regions $\Omega_i$ of the head are approximated with polygonal domains. On these domains, meshes are generated by using a triangular tessellation. The potential $\phi$ is approximated by a linear combination of expansion functions $\alpha_i \in X_\alpha$, i.e.

$$\phi \approx \sum_{j=1}^{N_\alpha} c_j \alpha_j \quad \text{for } \mathbf{r} \in \bigcup_i \Gamma_i, \tag{27}$$

where $c_j$ are the (unknown) expansion coefficients, and $N_\alpha = \dim X_\alpha$ is the dimensionality of the function space $X_\alpha$.

Commonly used functions are the piecewise constant functions (PCFs) $p_j$ (also referred to as patch functions) and Piecewise Linear Functions (PLFs) $\lambda_j$ (also referred to as linear Lagrangian or pyramid functions). These functions form the boundary element spaces span $\{p_j\}_{j=1}^{N_p} =: X_p$ and span $\{\lambda_j\}_{j=1}^{N_\lambda} =: X_\lambda$ and it holds that $X_p \subset H^{-1/2}$ and $X_\lambda \subset H^{1/2}$ (Steinbach, 2008). Both the PCFs and the PLFs form a partition of unity.

We denote the system matrix that stems from the discretization of an operator $X$ with the expansion functions $\alpha_j$ and testing functions $\beta_i$ as $\mathbf{X}_{\beta\alpha}$ with

$$\left[\mathbf{X}_{\beta\alpha}\right]_{ij} = \left(\beta_i, X\alpha_j\right)_{L^2}. \tag{28}$$

This notation is necessary, as some operators are discretized with different expansion and testing functions.

A standard discretization in literature for the adjoint double layer and double layer formulation is the one where PCFs are used as expansion and testing functions (Kybic et al., 2005; Stenroos and Sarvas, 2012). Using the notation of Eq. (28), such a discretization for the adjoint double layer formulation would read

$$\mathbf{I}_{k,\mathrm{pp}} - \mathbf{D}^*_{kl,\mathrm{pp}} = \mathbf{v}_{\mathrm{s1},k,\mathrm{p}} \quad k, l = 1, \ldots, N \tag{29}$$

where

$$\left[\mathbf{I}_{k,\mathrm{pp}}\right]_{ij} = \frac{\sigma_k + \sigma_{k+1}}{2(\sigma_{k+1} - \sigma_k)} \left(p_i^{(k)}, p_j^{(k)}\right)_{L^2(\Gamma_k)}, \tag{30}$$

$$\left[\mathbf{D}^*_{kl,\mathrm{pp}}\right]_{ij} = \left(p_i^{(k)}, \mathcal{D}^*_{kl} p_j^{(l)}\right)_{L^2(\Gamma_k)}, \tag{31}$$

$$\left[\mathbf{v}_{\mathrm{s1},k,\mathrm{p}}\right]_i = \left(p_i^{(k)}, \partial_{\hat{\mathbf{n}}} v_{\mathrm{s1}}\right)_{L^2(\Gamma_k)}, \tag{32}$$

and $q_{\Gamma_k} \approx \sum_i \left[\mathbf{q}_{k,\mathrm{p}}\right]_i p_i^{(k)}$.

The Petrov-Galerkin theory reviewed in the previous section, however, prohibits the use of PCFs as testing functions since on the right-hand-side we have the sum of the identity operator $\mathcal{I}$ and the adjoint double layer operator $\mathcal{D}^*$ terms. Since $\mathcal{I} : H^{-1/2} \to H^{-1/2}$, the entire right hand side is a (bijective) mapping from $H^{-1/2}$ to $H^{-1/2}$. The dual space of the range is $H^{1/2}$. As $X_p \not\subseteq H^{1/2}$, we cannot use PCFs as testing functions: they are not regular enough and thus the commonly used discretization of the operator in Eq. (29) is incompatible with the mapping $H^{-1/2} \to H^{-1/2}$ of the operator.

Also the classical discretization for the double layer formulation (Eq. (21)) is leveraging on PCF both as expansion and testing functions, i.e.

$$\mathbf{J}_{k,\mathrm{pp}} + \zeta_l \mathbf{D}_{kl,\mathrm{pp}} = \mathbf{v}_{\mathrm{s2},k,\mathrm{p}} \quad k, l = 1, \ldots, N \tag{33}$$

where

$$\left[\mathbf{J}_{k,\mathrm{pp}}\right]_{ij} = \frac{\sigma_k + \sigma_{k+1}}{2} \left(p_i^{(k)}, p_j^{(k)}\right)_{L^2(\Gamma_k)}, \tag{34}$$

$$\left[\mathbf{D}_{kl,\mathrm{pp}}\right]_{ij} = \left(p_i^{(k)}, \mathcal{D}_{kl} p_j^{(l)}\right)_{L^2(\Gamma_k)}, \tag{35}$$

$$\left[\mathbf{v}_{\mathrm{s2},k,\mathrm{p}}\right]_i = -\left(p_i^{(k)}, v_{\mathrm{s2}}\right)_{L^2(\Gamma)_k}, \tag{36}$$

and $\phi_{\Gamma_k} \approx \sum_i \left[\boldsymbol{\phi}_{k,\mathrm{p}}\right]_i p_i^{(k)}$ and $\zeta_l = -(\sigma_{l+1} - \sigma_l)$. For the double layer formulation Eq. (33), we find that the operator on the left-hand side is a mapping $H^{1/2} \to H^{1/2}$, since $\mathcal{I}$ maps from $H^{1/2}$ to $H^{1/2}$. The relationship $X_p \nsubseteq H^{1/2}$ implies that the use of PCFs as expansion functions is forbidden. More regular expansion functions must be used.

The standard discretization of the symmetric formulation reads (Kybic et al., 2005)

$$-\sigma_k \mathbf{N}_{kk-1,\lambda\lambda} + \mathbf{D}^*_{kk-1,\lambda p} + \eta_k \mathbf{N}_{kk,\lambda\lambda} - 2\mathbf{D}^*_{kk,\lambda p}$$
$$-\sigma_{k+1} \mathbf{N}_{kk+1,\lambda\lambda} + \mathbf{D}^*_{kk+1,\lambda p} = \mathbf{v}_{\text{sym},k,\lambda} \quad (37)$$

$$\mathbf{D}_{kk-1,p\lambda} - \sigma_k^{-1} \mathbf{S}_{kk-1,pp} - 2\mathbf{D}_{kk,p\lambda} + \theta_k \mathbf{S}_{kk,pp}$$
$$+ \mathbf{D}_{kk+1,p\lambda} - \sigma_{k+1}^{-1} \mathbf{S}_{kk+1,pp} = \mathbf{p}_{\text{sym},k,p} \quad (38)$$

where

$$\left[\mathbf{N}_{kl,\lambda\lambda}\right]_{ij} = \left(\lambda_i^{(k)}, \mathcal{N}_{kl} \lambda_j^{(l)}\right)_{L^2(\Gamma_k)}, \quad (39)$$

$$\left[\mathbf{S}_{kl,pp}\right]_{ij} = \left(p_i^{(k)}, \mathcal{S}_{kl} p_j^{(l)}\right)_{L^2(\Gamma_k)}, \quad (40)$$

$$\left[\mathbf{D}_{kl,p\lambda}\right]_{ij} = \left(p_i^{(k)}, \mathcal{D}_{kl} \lambda_j^{(l)}\right)_{L^2(\Gamma_k)}, \quad (41)$$

$$\left[\mathbf{D}^*_{kl,\lambda p}\right]_{ij} = \left(\lambda_i^{(k)}, \mathcal{D}^*_{kl} p_j^{(l)}\right)_{L^2(\Gamma_k)}, \quad (42)$$

$$\left[\mathbf{v}_{\text{sym},k,\lambda}\right]_i = \left(\lambda_i^{(k)}, \sigma_{k+1}^{-1}(v_{s,\Omega_{k+1}})|_{\Gamma_k} - \sigma_k^{-1}(v_{s,\Omega_k})|_{\Gamma_k}\right)_{L^2}, \quad (43)$$

$$\left[\mathbf{p}_{\text{sym},k,p}\right]_i = \left(p_i^{(k)}, (\partial_{\hat{\mathbf{n}}} v_{s,\Omega_{k+1}})|_{\Gamma_k} - (\partial_{\hat{\mathbf{n}}} v_{s,\Omega_k})|_{\Gamma_k}\right)_{L^2}, \quad (44)$$

with $\eta_i = \sigma_i + \sigma_{i+1}$, $\theta_i = \sigma_i^{-1} + \sigma_{i+1}^{-1}$, $\phi_{\Gamma_k} \approx \sum_i \left[\boldsymbol{\phi}_{k,\lambda}\right]_i \lambda_i^{(k)}$, and $d_{\Gamma_k} \approx \sum_i \left[\boldsymbol{\phi}_{k,p}\right]_i p_i^{(k)}$. In contrast to the classical discretizations of the double layer and adjoint double layer operators, the symmetric formulation in (Kybic et al., 2005) is discretized in a way that is completely conforming with respect to the fractional order Sobolev space mappings. The single layer operator $\mathcal{S}_{ij}$ maps from $H^{-1/2}$ to $H^{1/2}$ and thus the dual of its range is the space $H^{-1/2}$ allowing the use of PCFs as expansion and testing functions (see Eq. (43)). The hypersingular operator $\mathcal{N}_{ij}$ is discretized with PLFs as both expansion and testing functions, since the derivatives render the usage of PCFs impossible (see Eq. (37)). This is a conforming choice given that a lower regularity would not be allowed since $\mathcal{N}_{ij}$ is a mapping from $H^{1/2}$ to $H^{-1/2}$ and thus both expansion and testing functions should belong at least to $H^{1/2}$. Similar arguments can be used to show the conformity of the expansion and testing function choices for $\mathcal{D}_{ij}$ and $\mathcal{D}^*_{ij}$ in Eq. (41) and Eq. (42), respectively.

Furthermore, while the double layer and the adjoint double layer approaches give rise to a dense matrices, the matrices obtained from the symmetric approach are band diagonal. The entries of the matrix of the symmetric approach include only the interaction between adjacent compartments, which consequently reduces the computational cost.

Yet, the symmetric approach has two drawbacks: Its discretization gives rise to a matrix which is one and half times the size of the matrices of the previous two approaches and since the symmetric formulation operator is of the first kind, the resulting matrix is ill-conditioned (when the number of unknowns is increased by decreasing the average edge length $h$, the condition number grows unbounded (Steinbach and Wendland, 1998). We note that the double layer and adjoint double layer formulations are Fredholm integral equations of the second kind (Hackbusch, 2012). This kind of equations gives rise to well-conditioned systems.

Higher order functions could be used to solve the above mentioned problems in standard discretizations of the double and adjoint double layer operator. For example, given that the PLFs belong to the space $H^{1/2}$ and this space itself is a subset of the space $H^{-1/2}$, one could think of using PLFs as expansion and testing functions in Eq. (29) and Eq. (33). Although this leads to a conforming testing, it results in schemes that have either expansion or testing functions more regular than necessary, which can slow down convergence of the approximate solution to the exact solution as $h$ decreases; this effect is especially notable in the presence of irregular geometries such as highly realistic phantoms of the human head. Moreover, the usage of PLFs increases the computational burden: the handling of the singularity of the Green's function (Graglia, 1987, 1993; De Munck, 1992) in the integration routines becomes computationally more expensive and more difficult to implement.

Summarizing, the standard low order discretizations of the EEG forward problem do not comply with the EEG operators requirements in terms of regularity and can lead to erroneous solutions, while the symmetric formulation is correctly discretized but requires four times the memory space and it is an ill-conditioned

Table 1: The abbreviations and discretizations of the standard formulations.

| Label | Formulation | Function | | Drawbacks |
|---|---|---|---|---|
| | | Expansion | Testing | |
| **Standard Discretizations** | | | | |
| 1Aa | Adj. double layer | $p$ | $p$ | UR for testing |
| 1Ab | Adj. double layer | $\lambda$ | $\lambda$ | ER for expansion |
| 1Ba | Double layer | $p$ | $p$ | UR for expansion |
| 1Bb | Double layer | $\lambda$ | $\lambda$ | ER for testing |
| 1Bb-ISA | ISA | $\lambda$ | $\lambda$ | ER for testing |
| 1C | Symmetric | $p, \lambda$ | $p, \lambda$ | Double-sized and ill-conditioned system matrix |

ER : Excessive regularity

UR : Insufficient regularity

ISA : Isolated skull approach

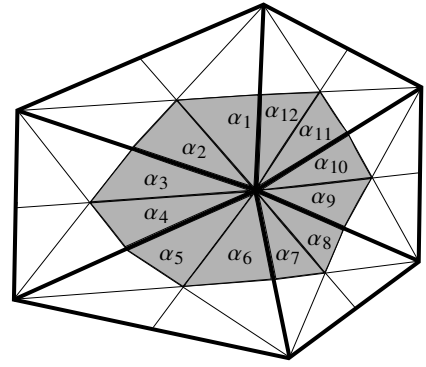

Figure 2: The standard (thick lines) and the barycentrically refined (thin lines) mesh. The grey area is a dual cell.

formulation. The problem could be ameliorated by higher order functions, but this increases the computational burden, and it is more complicated to implement. The above considerations are summarized in Tab. 1.

*1.3. New Mixed Discretized EEG Formulations*

A joint application of PCFs and PLFs is not possible without getting a rectangular matrix – owing to the fact that the number of PCFs $N_\mathrm{p}$ equals the number of cells of the mesh $N_\mathrm{Cells}$, while the number of PLFs $N_\lambda$ equals the number of inner nodes $N_\mathrm{Nodes}$ of the mesh.

We must use a new set of PCFs and PLFs whose span has the right dimension. Such functions can be constructed by using the dual mesh, that is, the mesh where the nodes of the original mesh become cells and vice versa. In the following, we refer to these functions as Dual Piecewise Constant Functions (DPCFs) $\widetilde{p}_i$ and Dual Piecewise Linear Functions (DPLFs) $\widetilde{\lambda}_i$, and they form the spaces $X_{\widetilde{\mathrm{p}}}$ and $X_{\widetilde{\lambda}}$.

The dual mesh is obtained by barycentrically refining a standard triangular mesh: each triangle is split into six sub-triangles by connecting the midpoints of each edge with the opposite node. This is shown in Fig. 2, where the original cells are those with bold edges. The union of the greyed cells around the center node of this figure form the dual cell.

In the following, a mixed discretization scheme for the EEG based forward problem is proposed. The operators are discretized and tested in a conforming way with respect to Sobolev space mapping properties. Mixed and conforming discretization techniques were introduced by (Cools et al., 2009) in the context of full wave solutions of scattering problems. These discretizations make use of suitably chosen dual functions. There are several ways to define these dual functions. In this work we adopt the functions proposed in (Buffa and Christiansen, 2007) in the context of Calderon preconditioning of scattering problems.

*1.3.1. Dual Functions*

The definition of the DPCFs is simple. The support of the DPCF $\widetilde{p}_j$ is given by the cells on the barycentrically refined mesh that are attached to the $j$th node (Fig. 2 shows the support of a DPCF). When $\boldsymbol{r}$ is in the support of $\widetilde{p}_j$, the function value $\widetilde{p}_j(\boldsymbol{r}) = 1$ and is zero when $\boldsymbol{r}$ is not in the support of $\widetilde{p}_j$. Let $p_i^\mathrm{bar}$ be the standard PCFs defined on the barycentrically refined mesh. For the example given in Fig. 2, we find

$$\widetilde{p}_j = \sum_{i=1}^{12} \alpha_i p_i^\mathrm{bar}, \qquad (45)$$

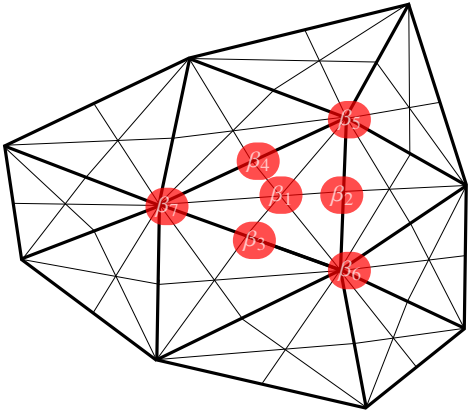

Figure 3: Transformation matrix coefficient

where $\alpha_i = 1$ for all $i = 1, \ldots, 12$.

The function $\widetilde{\lambda}_j$ is attached to the cell $j$ and it can be represented as a linear combination of seven standard PLFs $\lambda_j^{\text{bar}}$ defined on the barycentrically refined mesh. Fig. 3 shows the general case, where the nodes of the seven relevant PLFs are labeled with the coefficients $\beta_i$ that are chosen such that

$$\widetilde{\lambda}_j = \sum_{i=1}^{7} \beta_i \lambda_i^{\text{bar}}. \qquad (46)$$

For the first coefficient, which is associated with the center PLF, we always have $\beta_1 = 1$, while for the next three coefficients, which are associated with the PLFs defined on the midpoints of the edges of the primal cell, we always have $\beta_i = 1/2$ with $i = 2, \ldots, 4$. The last three coefficients $\beta_5$, $\beta_6$ and $\beta_7$, which are associated with PLFs defined on nodes of the primal cell, their weights are given by $1/N_{\text{Cells},i}$ with $i = 5, \ldots, 7$ and $N_{\text{Cells},i}$ being the number of cells of the primal mesh that are attached to the respective node. In the example given, the coefficients are $\alpha_5 = 1/5$, $\alpha_6 = 1/6$, and $\alpha_7 = 1/6$. It can be shown that these DPLFs form a partition of unity (Buffa and Christiansen, 2007). Fig. 4 visualizes an example of a DPLF function.

### 1.3.2. New Formulations

Following the considerations of the previous sections, we propose new mixed discretization strategies for the adjoint double layer, double layer and symmetric approaches. For the adjoint double layer approach, we use standard PCFs as expansion and DPLFs as testing functions. This results in

$$\begin{bmatrix} I_{1,\widetilde{\lambda}\text{p}} - D^*_{11,\widetilde{\lambda}\text{p}} & -D^*_{12,\widetilde{\lambda}\text{p}} & -D^*_{13,\widetilde{\lambda}\text{p}} \\ -D^*_{21,\widetilde{\lambda}\text{p}} & I_{2,\widetilde{\lambda}\text{p}} - D^*_{22,\widetilde{\lambda}\text{p}} & -D^*_{23,\widetilde{\lambda}\text{p}} \\ -D^*_{31,\widetilde{\lambda}\text{p}} & -D^*_{32,\widetilde{\lambda}\text{p}} & I_{3,\widetilde{\lambda}\text{p}} - D^*_{33,\widetilde{\lambda}\text{p}} \end{bmatrix} \cdot \underbrace{\begin{bmatrix} q_{1,\text{p}} \\ q_{2,\text{p}} \\ q_{3,\text{p}} \end{bmatrix}}_{q_{\text{p}}} = \begin{bmatrix} v_{\text{s1},1,\widetilde{\lambda}} \\ v_{\text{s1},2,\widetilde{\lambda}} \\ v_{\text{s1},3,\widetilde{\lambda}} \end{bmatrix}, \qquad (47)$$

where

$$\left[ I_{k,\widetilde{\lambda}\text{p}} \right]_{ij} = \frac{\sigma_k + \sigma_{k+1}}{2(\sigma_{k+1} - \sigma_k)} \left( \widetilde{\lambda}_i^{(k)}, p_j^{(k)} \right)_{L^2(\Gamma)}, \qquad (48)$$

$$\left[ D^*_{kl,\widetilde{\lambda}\text{p}} \right]_{ij} = \left( \widetilde{\lambda}_i^{(k)}, \mathcal{D}^*_{kl} p_j^{(l)} \right)_{L^2(\Gamma)}, \qquad (49)$$

$$\left[ v_{\text{s1},k,\widetilde{\lambda}} \right]_i = \left( \widetilde{\lambda}_i^{(k)}, \partial_{\hat{\boldsymbol{n}}} v_{\text{s1}} \right)_{L^2(\Gamma)}, \qquad (50)$$

and $q_{\Gamma_k} \approx \sum_i \left[ q_{k,\text{p}} \right]_i p_i^{(k)}$.

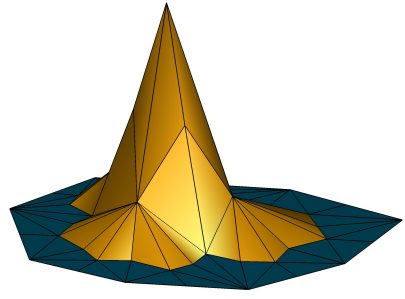

Figure 4: An example of a typical BC piecewise linear function.

For the double layer approach, on the other hand, we use standard PLFs as expansion and DPCFs as testing functions. The system we have to solve is

$$\begin{bmatrix} J_{1,\widetilde{\text{p}}\lambda} + \zeta_1 D_{11,\widetilde{\text{p}}\lambda} & \zeta_2 D_{12,\widetilde{\text{p}}\lambda} & \zeta_3 D_{13,\widetilde{\text{p}}\lambda} \\ \zeta_1 D_{21,\widetilde{\text{p}}\lambda} & J_{2,\widetilde{\text{p}}\lambda} + \zeta_2 D_{22,\widetilde{\text{p}}\lambda} & \zeta_3 D_{23,\widetilde{\text{p}}\lambda} \\ \zeta_1 D_{31,\widetilde{\text{p}}\lambda} & \zeta_2 D_{32,\widetilde{\text{p}}\lambda} & I_{3,\widetilde{\text{p}}\lambda} + \zeta_3 D_{33,\widetilde{\text{p}}\lambda} \end{bmatrix} \cdot \underbrace{\begin{bmatrix} \boldsymbol{\phi}_{1,\lambda} \\ \boldsymbol{\phi}_{2,\lambda} \\ \boldsymbol{\phi}_{3,\lambda} \end{bmatrix}}_{\boldsymbol{\phi}_{\text{p}}} = \begin{bmatrix} v_{\text{s2},1,\widetilde{\text{p}}} \\ v_{\text{s2},2,\widetilde{\text{p}}} \\ v_{\text{s2},3,\widetilde{\text{p}}} \end{bmatrix}, \qquad (52)$$

Table 2: The abbreviations and discretizations of the new formulations.

| Label | Formulation | Function | |
|---|---|---|---|
| | | Expansion | Testing |
| **Mixed Discretizations (This work)** | | | |
| 2A | Adjoint double layer | $p$ | $\widetilde{\lambda}$ |
| 2B | Double layer | $\lambda$ | $\widetilde{p}$ |
| 2B-ISA | Isolated skull approach | $\lambda$ | $\widetilde{p}$ |
| 2C | New Symmetric | $p, \lambda$ | $\widetilde{p}, \lambda$ |

where

$$\left[\boldsymbol{J}_{k,\widetilde{p}\lambda}\right]_{ij} = \frac{\sigma_k + \sigma_{k+1}}{2} \left(\widetilde{p}_i^{(k)}, \lambda_j^{(k)}\right)_{L^2(\Gamma)}, \quad (53)$$

$$\left[\boldsymbol{D}_{kl,\widetilde{p}\lambda}\right]_{ij} = \left(\widetilde{p}_i^{(k)}, \mathcal{D}_{kl}\, \lambda_j^{(l)}\right)_{L^2(\Gamma)}, \quad (54)$$

$$\left[\boldsymbol{v}_{s2,k,\widetilde{p}}\right]_{i} = -\left(\widetilde{p}_i^{(k)}, v_{s2}\right)_{L^2(\Gamma)}, \quad (55)$$

and $\phi_{\Gamma_k} \approx \sum_i \left[\boldsymbol{\phi}_{k,\lambda}\right]_i \lambda_i^{(k)}$ and $\zeta_l = -(\sigma_{l+1} - \sigma_l)$.

For the sake of completeness, we also consider a mixed discretization for the symmetric formulation. When a mixed discretization is applied to Eq. (23) and Eq. (24), we obtain a rectangular system matrix. To obtain a square system matrix, we propose a slightly modified symmetric formulation given by

$$\begin{aligned}
\left(v_{s,\Omega_{i+1}}\right)\big|_{\Gamma_i} &- \left(v_{s,\Omega_i}\right)\big|_{\Gamma_i} \\
&= \sigma_i \mathcal{D}_{i,i-1}\phi_{\Gamma_{i-1}} - (\sigma_i + \sigma_{i+1}) \mathcal{D}_{ii}\phi_{\Gamma_i} \\
&\quad - (\sigma_i - \sigma_{i+1})\phi_{\Gamma_i} + \sigma_{i+1}\mathcal{D}_{i,i+1}\phi_{\Gamma_{i+1}} \\
&\quad - \mathcal{S}_{i,i-1}d_{\Gamma_{i-1}} + 2\mathcal{S}_{ii}d_{\Gamma_i} + \mathcal{S}_{i,i+1}d_{\Gamma_{i+1}} \quad (56)
\end{aligned}$$

and

$$\begin{aligned}
\left(\sigma_i^{-1}\partial_{\hat{\boldsymbol{n}}} v_{s,\Omega_{i+1}}\right)\big|_{\Gamma_i} &- \left(\sigma_{i+1}^{-1}\partial_{\hat{\boldsymbol{n}}} v_{s,\Omega_i}\right)\big|_{\Gamma_i} \\
&= \mathcal{N}_{i,i-1}\phi_{\Gamma_{i-1}} - 2\mathcal{N}_{ii}\phi_{\Gamma_i} + \mathcal{N}_{i,i+1}\phi_{\Gamma_{i+1}} \\
&\quad - \sigma_i^{-1}\mathcal{D}_{i,i-1}^* d_{\Gamma_{i-1}} + \left(\sigma_i^{-1} + \sigma_{i+1}^{-1}\right) \mathcal{D}_{ii}^* d_{\Gamma_i} \\
&\quad - \left(\sigma_i^{-1} - \sigma_{i+1}^{-1}\right) d_{\Gamma_i} - \sigma_{i+1}^{-1}\mathcal{D}_{i,i+1}^* d_{\Gamma_{i+1}} \quad (57)
\end{aligned}$$

In the case of three layers, the explicit expression of the resulting linear system is given in Eq. (51), where we use

$$\left[\boldsymbol{N}_{kl,\widetilde{\lambda}\lambda}\right]_{ij} = \left(\widetilde{\lambda}_i^{(k)}, \mathcal{N}_{kl}\lambda_j^{(l)}\right)_{L^2(\Gamma_k)}, \quad (58)$$

$$\left[\boldsymbol{S}_{kl,\widetilde{p}p}\right]_{ij} = \left(\widetilde{p}_i^{(k)}, \mathcal{S}_{kl}p_j^{(l)}\right)_{L^2(\Gamma_k)}, \quad (59)$$

$$\left[\boldsymbol{D}_{kl,\widetilde{p}\lambda}\right]_{ij} = \left(\widetilde{p}_i^{(k)}, \mathcal{D}_{kl}\lambda_j^{(l)}\right)_{L^2(\Gamma_k)}, \quad (60)$$

$$\left[\boldsymbol{D}_{kl,\widetilde{\lambda}p}^*\right]_{ij} = \left(\widetilde{\lambda}_i^{(k)}, \mathcal{D}_{kl}^*p_j^{(l)}\right)_{L^2(\Gamma_k)}, \quad (61)$$

$$\left[\boldsymbol{v}_{\text{sym},k,\widetilde{p}}\right]_{i} = \left(\widetilde{p}_i^{(k)}, (v_{s,\Omega_{k+1}})\big|_{\Gamma_k} - (v_{s,\Omega_k})\big|_{\Gamma_k}\right)_{L^2},$$

$$\left[\boldsymbol{p}_{\text{sym},k,\widetilde{\lambda}}\right]_{i} = \left(\widetilde{\lambda}_i^{(k)}, \left(\sigma_k^{-1}\partial_{\hat{\boldsymbol{n}}} v_{s,\Omega_{k+1}}\right)\big|_{\Gamma_k} - \left(\sigma_{k+1}^{-1}\partial_{\hat{\boldsymbol{n}}} v_{s,\Omega_k}\right)\big|_{\Gamma_k}\right)_{L^2},$$

with coefficients

$$\alpha_i = (\sigma_i + \sigma_{i+1}), \qquad \beta_i = (\sigma_i - \sigma_{i+1})/2,$$
$$\gamma_i = \left(\sigma_i^{-1} + \sigma_{i+1}^{-1}\right), \qquad \delta_i = \left(\sigma_{i+1}^{-1} - \sigma_i^{-1}\right)/2,.$$

An important point to note is that the BEM equations mentioned above correspond to the interior Neumann problem. Therefore, the potential is only determined up to an additive constant. This gives rise to singular matrices. Hence, an additional condition is needed to solve these systems. The most commonly used one is $\int_\Gamma \phi \mathrm{d}S(\boldsymbol{r}) = 0$ which corresponds to a surface potential of zero mean. We do not further detail this point given that this is a standard procedure which is widely documented (Rahola and Tissari, 2002; Chan, 1984).

$$\begin{bmatrix}
\beta_1 \boldsymbol{I}_{\widetilde{p}\lambda} + \alpha_1 \boldsymbol{D}_{11,\widetilde{p}\lambda} & -2\boldsymbol{S}_{11,\widetilde{p}p} & -\sigma_2 \boldsymbol{D}_{12,\widetilde{p}\lambda} & \boldsymbol{S}_{12,\widetilde{p}p} & \boldsymbol{0} \\
-2\boldsymbol{N}_{11,\widetilde{\lambda}\lambda} & \delta_1 \boldsymbol{I}_{\widetilde{\lambda}p} + \gamma_1 \boldsymbol{D}_{11,\widetilde{\lambda}p}^* & \boldsymbol{N}_{12,\widetilde{\lambda}\lambda} & -\sigma_2^{-1}\boldsymbol{D}_{12,\widetilde{\lambda}p}^* & \boldsymbol{0} \\
-\sigma_2 \boldsymbol{D}_{21,\widetilde{p}\lambda} & \boldsymbol{S}_{21,\widetilde{p}p} & \beta_2 \boldsymbol{I}_{\widetilde{p}\lambda} + \alpha_2 \boldsymbol{D}_{22,\widetilde{p}\lambda} & -2\boldsymbol{S}_{22,\widetilde{p}p} & -\sigma_3 \boldsymbol{D}_{23,\widetilde{p}\lambda} \\
\boldsymbol{N}_{21,\widetilde{\lambda}\lambda} & -\sigma_2^{-1}\boldsymbol{D}_{21,\widetilde{\lambda}p}^* & -2\boldsymbol{N}_{22,\widetilde{\lambda}\lambda} & \delta_2 \boldsymbol{I}_{\widetilde{\lambda}p} + \gamma_2 \boldsymbol{D}_{22,\widetilde{\lambda}p}^* & \boldsymbol{N}_{23,\widetilde{\lambda}\lambda} \\
\boldsymbol{0} & \boldsymbol{0} & -\sigma_3 \boldsymbol{D}_{32,\widetilde{p}\lambda} & \boldsymbol{S}_{32,\widetilde{p}p} & \beta_3 \boldsymbol{I}_{\widetilde{p}\lambda} + \alpha_3 \boldsymbol{D}_{33,\widetilde{p}\lambda}
\end{bmatrix}
\begin{bmatrix} \boldsymbol{\phi}_{1,\lambda} \\ \boldsymbol{d}_{1,p} \\ \boldsymbol{\phi}_{2,\lambda} \\ \boldsymbol{d}_{2,p} \\ \boldsymbol{\phi}_{3,\lambda} \end{bmatrix}
=
\begin{bmatrix} \boldsymbol{v}_{\text{sym},1,\widetilde{p}} \\ \boldsymbol{p}_{\text{sym},1,\widetilde{\lambda}} \\ \boldsymbol{v}_{\text{sym},2,\widetilde{p}} \\ \boldsymbol{p}_{\text{sym},2,\widetilde{\lambda}} \\ \boldsymbol{v}_{\text{sym},3,\widetilde{p}} \end{bmatrix}, \quad (51)$$

The integrals found in the operators $\mathcal{D}$ and $\mathcal{D}*$ are computed numerically using Gauss quadrature (Dunavant, 1985) with seven points when the expansion and testing triangle are far from each other. When the expansion and testing triangles are too close such that the singular kernel cannot be numerically integrated accurately enough, we employ a singularity extraction technique (Graglia, 1993) for the inner integrations. For the integration of the $\mathcal{N}$ operator, we follow a standard approach by using its weak formulation such that the derivatives are placed on the testing and expansion functions (Steinbach, 2008). The identity operators are always computed analytically.

## 2. Results

To show the practical impact of our newly proposed schemes, we compared them with some of the most common BEM formulations. A first set of comparisons has been obtained on the traditional multi-layered spherical model (De Munck, 1988; Munck and Peters, 1993; Zhang, 1995). The reason for this choice is that the analytic solution available for such a model provides a rigorous reference for solidly assessing the numerical performance of all different formulations. This set of comparisons is then complemented by a second one where we show that our new methods are naturally applicable to MRI-obtained models and that also in this case their performance compares quite favourably with the existing techniques.

### 2.1. Numerical Experiments on a Layered Spherical Head Model

The radii of the concentric spheres of the model (Fig. 5) are 0.87 dm, 0.92 dm, and 1 dm. Each sphere is triangulated with an average edge length $h = 0.17$ dm resulting in 700 elements on first layer, 796 elements on the second layer and 920 elements on the third layer. The triangulation is uniform delivering a mesh of approximately equally sized and shaped elements. The corresponding (normalized) conductivities are $\sigma_1 = 1$, $\sigma_2 = 1/15$, and $\sigma_3 = 1$, which we abbreviate with the ratio 1:1/15:1 (Oostendorp et al., 2000). The source is modeled by a current dipole of magnitude one and of orientation $\begin{bmatrix} 1 & 0 & 1 \end{bmatrix}^\mathsf{T}$.

In Tabs. 1 and 2, we summarize the different discretization strategies and, accordingly, we reference these strategies in the legends

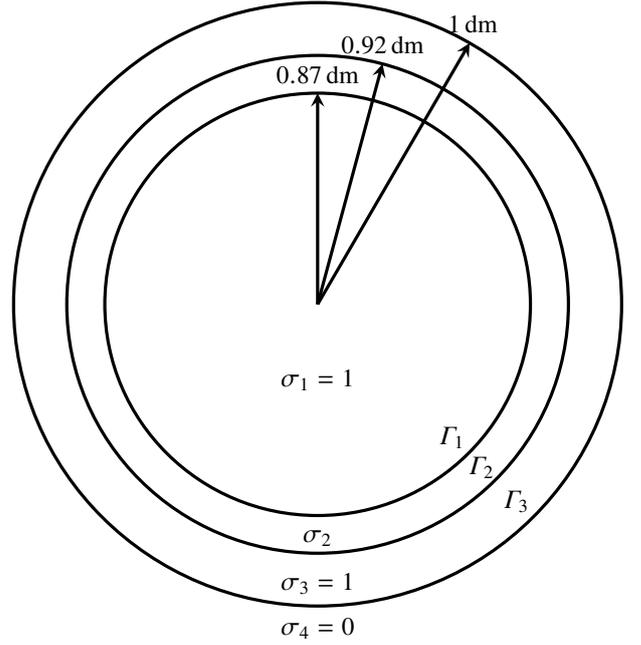

Figure 5: The multi-layered spherical structure that is used for the evaluation of the different discretizations. Note, that the conductivity $\sigma_2$ was varied in parts of the simulations.

of the plots.

In this scenario, we conducted three simulations where the position of the dipole was the varying factor. In the first experiment (Fig. 6), the dipole was moved along the line between the center of the spherical model and a triangle vertex located on the innermost sphere, in the second experiment (Fig. 7), the dipole was moved along the line between a centroid of a triangle laying on the innermost sphere and the center of the spheres. Finally, in the third experiment (Fig. 8), the dipole was moved along a line between the center of the spherical model and a point that does not correspond to centroid nor to a node. All figures 6 to 8 show, for different formulations, the relative error with respect to the analytic solution as a function of the source dipole's radial distance from the center of the spheres. We see that the relative error depends on the position of current source with respect to the nearest vertex. We also see that, in general, the relative error increases when the dipole approaches the surface $\Gamma_1$. However, the formulations properly discretized are less affected by this phenomenon than other formulations. The dual adjoint double layer and the symmetric formulations presents more accurate results in all cases.

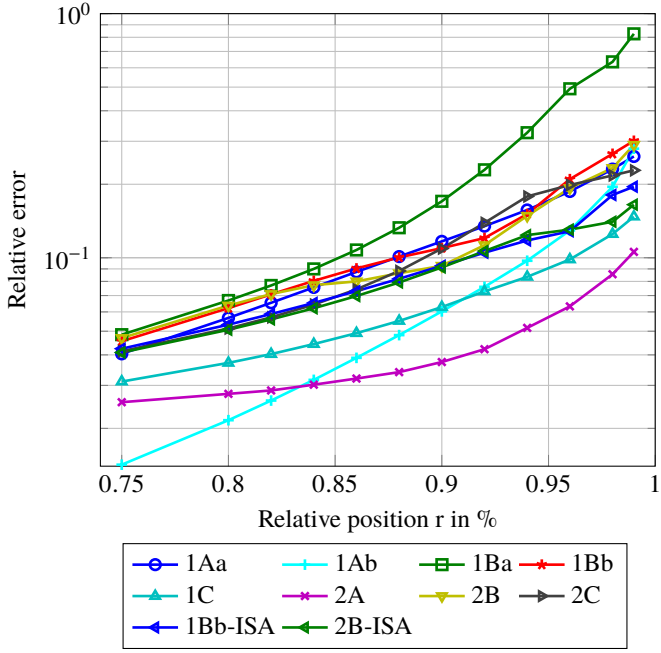
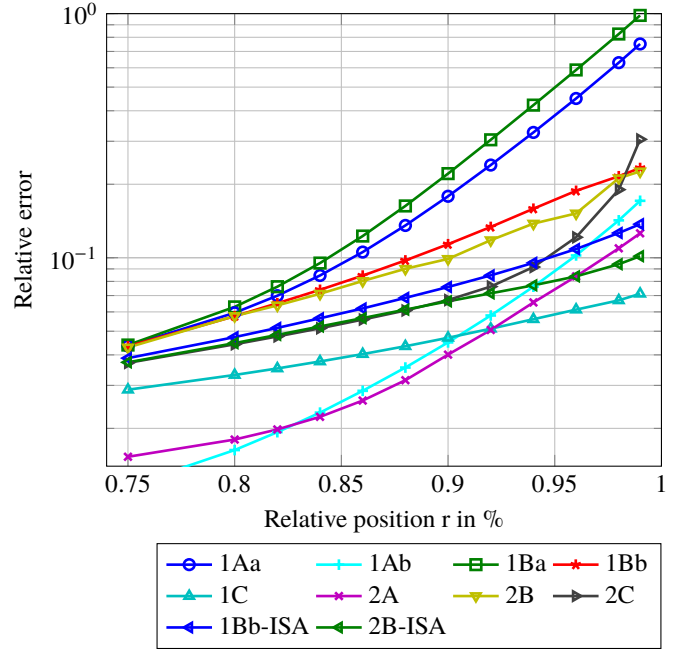

Figure 6: relative error as a function of the dipole position inside the sphere $\Gamma_1$ with the conductivity $\sigma_2 = 1/15$ S m$^{-1}$. The dipole is under a vertex.

Figure 7: relative error as a function of the dipole position inside the sphere $\Gamma_1$ with the conductivity $\sigma_2 = 1/15$ S m$^{-1}$. The dipole is under a centroid.

It is well known that the BEM formulations depend on the quality of mesh that discretizes the computational domain. Therefore we have runned another set of simulations where the spherical model was discretized non uniformly. This means that the triangle's nodes are arbitrarily located over the three spheres and the number of cells connected to a particular node vary from node to node. The total number of triangles on each layer was $N = 900$. After computing the scalp potential for different current source positions, we reported the relative error in fig. 9, where an analytical solution was used as reference. We observe that the behavior of the different formulations is similar to the previous case in which the dual adjoint double layer provides the highest accuracy.

The behavior as well as the accuracy of the different BEM formulations depend on the conductivity ratio of the different compartments of the head. This aspect has been investigated and the results are shown in Fig. 10 displaying the relative error as a function of the electric resistivity $1/\sigma_2$ when the dipole is positioned at 0.83 dm along the $x$-axis. We observe that the standard non-conforming discretizations of double layer and adjoint double layer operator (1Aa and 1Ba) are more strongly impacted by the change of the conductivity $\sigma_2$ than the other formulations. In addition, we observe that the mixed adjoint double layer formulation is superior when the normalized electric resistivity $1/\sigma_2$ is below 45. The isolated skull approach (ISA) is often used in literature in the presence of a layer of low conductivity to mitigate its impact on the solution error Hamalainen and Sarvas (1989); Meijs et al. (1989); Gençer and Akalin-Acar (2005). Figure 10 shows also the impact of the use of the ISA technique on both standard formulations and on the mixed discretized schemes that we propose in this work. It is found that the benefits of the ISA and of the discretization technique we propose are cumulative and that the two techniques can be perfectly used together.

Functional assessment and cerebral diagnostic of brain activity of newborns has motivated an increased interest in neonatal EEG source localization (Patrizi et al., 2003; Watanabe et al., 1999). The skull conductivity of infants differs from the skull conductivity of adults (Ernst et al., 2011) and a head model of one compartment is often used to obtain a reliable source localization (Odabaee et al., 2012; Silau et al., 1994; Odabaee et al., 2014). Thus we used a single sphere with homogeneous conductivity to model this scenario (Despotovic et al., 2013). Fig. 11 shows the relative error of the different EEG forward problem methods for this scenario. The mixed adjoint double layer approach shows the best performance compared with any other formulation.

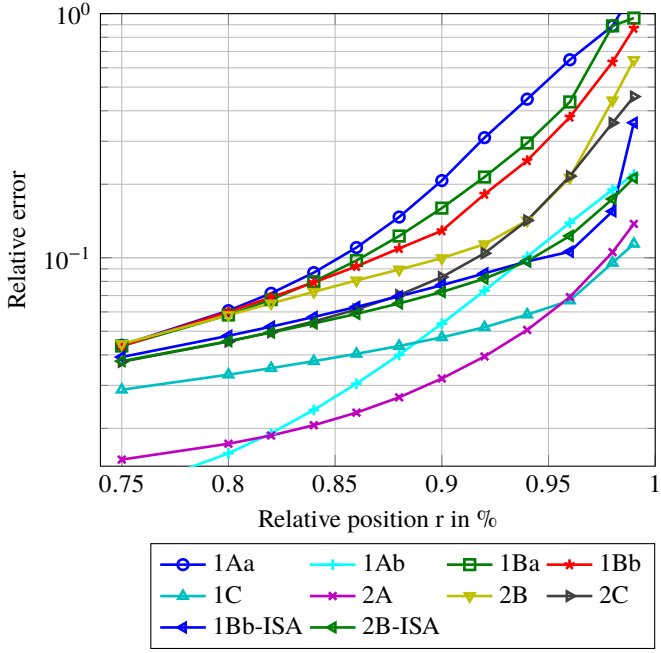
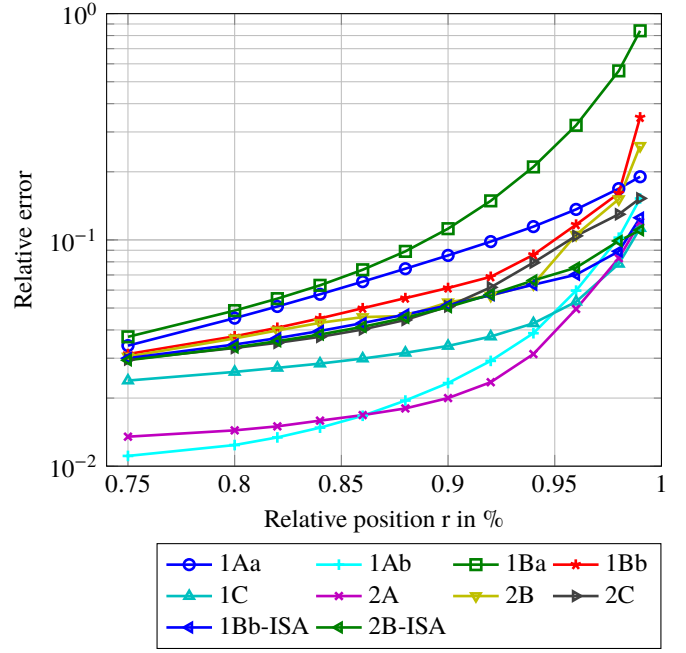

Figure 8: relative error as a function of the dipole position inside the sphere $\Gamma_1$ with the conductivity $\sigma_2 = 1/15\,\mathrm{S\,m^{-1}}$. The dipole is not under a vertex nor under a centroid.

Figure 9: relative error as a function of the dipole position inside the sphere $\Gamma_1$ with the conductivity $\sigma_2 = 1/15\,\mathrm{S\,m^{-1}}$. Nonuniform meshing.

*2.2. Numerical experiments on an MRI-obtained head model*

Although spherical models (and associated analytical solutions) are fundamental for a robust assessment of any newly proposed forward solution strategy, it is of fundamental importance to show the applicability and performance of the technique proposed here on realistic MRI-obtained head models. These models allow for an individual-based head model to be used in solving the forward problem and translates in more precise source localization Huang et al. (2015). Different well-established methods exist in the literature for extracting the cerebral interfaces and are available in several commercial and academic packages including Curry, ScanIP, ASA, BESA, FieldTrip Oostenveld et al. (2010), FMRIB Smith (2002), FreeSurfer Fischl et al. (2004), BrainVISA Geffroy et al. (2011), BrainSuite Shattuck and Leahy (2002), 3D Slicer Fedorov et al. (2012), and BrainVoyager Goebel (2012). In our numerical experiment we have leveraged on BrainSuite to obtain an automatic segmentation of the brain, CSF, skull and scalp. The MRI images used here are a T1-weighted scans of $256 \times 256 \times 256$ cubic voxels (refer to Fig. 12). After following the standard pipeline to reconstruct the surfaces of head tissues, the obtained model is made of 97 338, 50 398 and 40 634 cells for the brain, the skull and the scalp respectively. The conductivity of the brain, CSF, skull and the scalp is $\sigma_1 = 1$, $\sigma_2 = 5.42$, $\sigma_2 = 1/15$, and $\sigma_3 = 1$ respectively.

After the forward problem has been solved, the resulting surface potential can be visualized in Fig. 13. Since the exact solution is unavailable in this case, FEM served as a reference. It was solved on refined model having 6 million tetrahedrons and the relative error was calculated at the position of 128 electrodes positioned according to an EGI system (see Fig. 14). The results of this benchmarking can be seen in Fig. 15, where a dipolar source was moved from the center of the right hemisphere along the z axis and the relative error with respect to the reference solution has been computed for all approaches presented in the previous section. We observe that the dual adjoint double layer and the symmetric formulations delivers the most accurate results.

Tab. 3 summarizes the measurement of the memory and computational time to solve the forward problem. Timing measurements concerns only the adjoint double layer and the double layer formulations with different discretization schemes. The machine used is a Dell server R920 working under windows system. From the table we conclude that proposed mixed discretization is not more computational demanding while complying with the operators mapping properties.

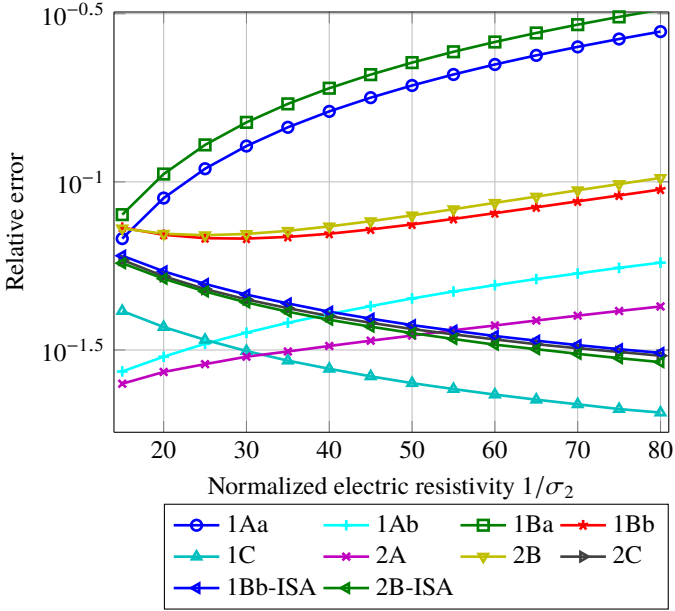

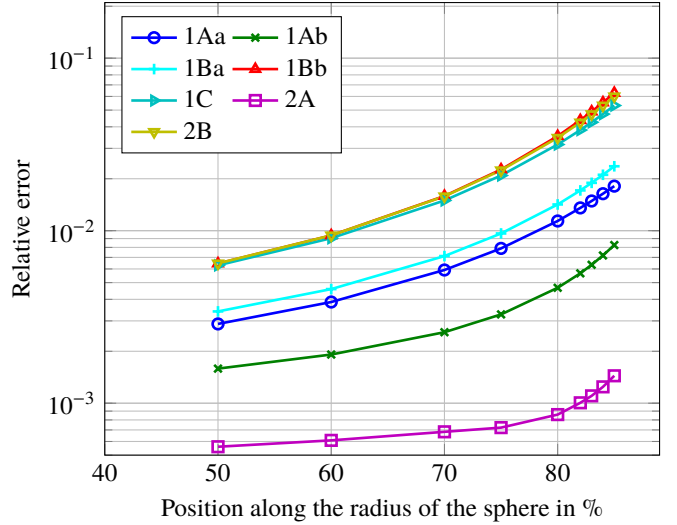

Figure 11: Single compartment sphere: relative error as a function of the dipole position.

Figure 10: relative error as a function of the normalized electric resistivity $1/\sigma_2$ when the dipole is positioned at 0.83 dm along the $x$-axis.

## 3. Discussion and conclusions

Numerical results confirm that classical, commonly used, discretizations of the integral operators $\mathcal{D}$ and $\mathcal{D}^*$ result in substantially lower level of accuracy than those achievable with their mixed discretized counterparts we propose here (Figs. 6 to 9). If we then look for the best possible forward problem formulations across all different operators, depending on the scenario (position of the dipole, conductivity), either the mixed discretization or the piecewise linear PLFs discretization of the adjoint double layer operator (2A and 1Ab) provide the overall best accuracy. The reader should remember however, that the PLF discretization always comes at the price of an increase in the computational burden. Moreover, every time the dipolar brain source comes close to a brain layer interface, we find that the mixed discretization yields always the best accuracy.

We varied the conductivity over a wide range of values. In recent reports the conductivity ratio of brain, skull, and scalp is estimated from in-vivo measurements to be between 1:1/15:1 and 1:1/25:1, and the ratio 1:1/80:1 arising from in vitro measurements and which until recently has been used in neuroimaging applications is currently questioned (Gonçalves et al., 2003; Zhang et al., 2006; Oostendorp et al., 2000; Clerc et al., 2005). We have observed that for the relevant range of in vivo measured conductivities, the mixed adjoint double layer formulation yields the highest accuracy compared with other methods. This is also the case for the scenario where a head of a newborn is modelled (see Fig. 11): the mixed discretization of the adjoint double layer formulation obtains the highest accuracy.

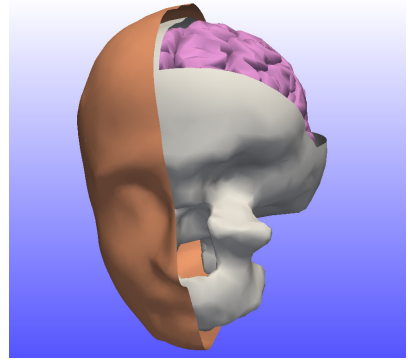

Figure 12: MRI-based head model: Scalp, skull, brain, and CSF.

From our experiments it is evident that also the symmetric formulation performs quite well when the dipole is near the surface (see Figs. 6 to 9). This is consistent with the theoretical consider-ation that the symmetric formulation is a conformingly discretized scheme. However, it should be noted that the symmetric formulation comes at the cost of a considerable computational burden. The dimension of the system matrix is one and a half times the size of the matrices of the mixed discretization. Thus the symmetric formulation is comparatively less competitive in the source localization process, where the propagation model needs to be calculated many times.

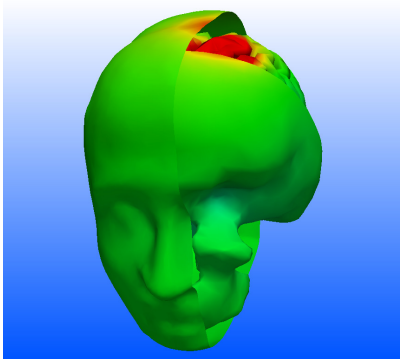

Figure 13: MRI-based head model: surface potential

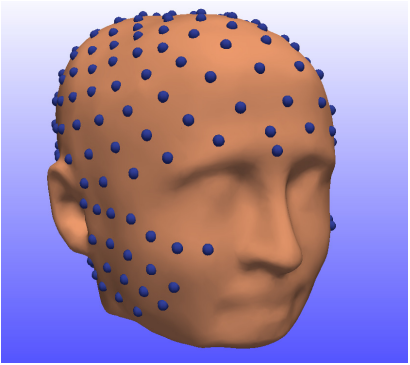

Figure 14: MRI-based head model: electrodes' positions

For the sake of completeness, we also presented a mixed discretization of the symmetric formulation. No advantages were observed, since the original symmetric formulation was already discretized conformingly, and the obtained system is plagued with the same drawbacks as the original symmetric formulation. We note also that the application of the isolated skull approach improves the accuracy of the solution, especially for shallow dipoles. The improvement, however, comes at the cost of some additional computations.

In conclusion, the numerical results confirm what can be theo-retically expected, i.e. that mixed and conforming discretizations provide formulations with a higher level of accuracy of their standard counterparts. Moreover the adjoint double layer operator provides often the global optimum across all operators and discretizations. Taking into account that the computational costs of the mixed discretization is lower than those of higher order alternatives, the discretizations schemes proposed here can be a very competitive new option among all EEG forward formulations.

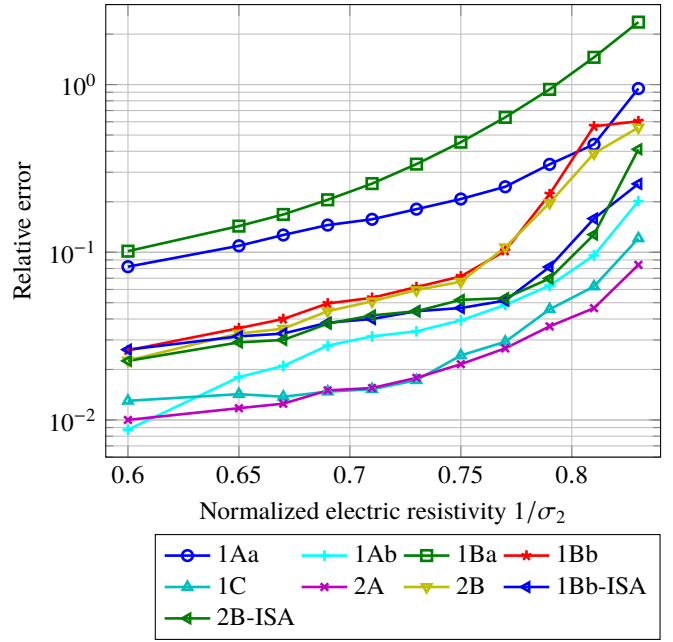

Figure 15: Head: Relative error as a function of the normalized electric resistivity $1/\sigma_2$ when the dipole is positioned at 0.83 dm along the $x$-axis.

## Acknowledgement

This work was supported in part by the Agence Nationale de la Recherche under Project FASTEEG-ANR-12-JS09-0010, in part by the European Union Project NEUROIMAGEEG, and in part by the HPC resources through the GENCI-TGCC Project under Grant 2015-gen6944.

## Appendix A. Relevant Sobolev Spaces

This appendix contains the definitions of the Sobolev spaces used in this paper. The presentation is concise for the sake of brevity; the interested reader could refer to (Tartar, 2007) for further details on the topic.

Table 3: Memory and time consumption for the double layer and adjoint double layer discretizations.

| Function | | CPU | Memory | Time | |
|---|---|---|---|---|---|
| Expansion | Testing | | | CG | Total |
| | | (#) | (GB) | (s) | (s) |
| **Adjoint Double Layer Discretizations** | | | | | |
| $p$ | $p$ | 54 | 642 | 350 | 19 790 |
| $p$ | $\widetilde{\lambda}$ | 54 | 642 | 408 | 25 786 |
| $\lambda$ | $\lambda$ | 54 | 160 | 103 | 37 672 |
| **Double Layer Discretizations** | | | | | |
| $p$ | $p$ | 54 | 642 | 340 | 18 486 |
| $\lambda$ | $\widetilde{p}$ | 54 | 160 | 120 | 24 210 |
| $\lambda$ | $\lambda$ | 54 | 160 | 100 | 35 321 |

The Sobolev space $H^1(\Omega)$ is defined as

$$H^1(\Omega) = \left\{ f : \Omega \to \mathbb{R} \mid f \in L^2(\Omega) \wedge \nabla f \in \left(L^2(\Omega)\right)^3 \right\}.$$

The space $L^2$ is the set of all equivalence classes of functions that are square integrable in the Lebesgue sense:

$$L^2(\Omega) = \left\{ f : \Omega \to \mathbb{R} \mid \|f\|_{L^2(\Omega)} < \infty \right\}, \quad \text{(A.1)}$$

with the $L^2$-norm defined as

$$\|f\|_{L^2(\Omega)} = \left( \int_\Omega |f(\boldsymbol{r})|^2 d\boldsymbol{r} \right)^{1/2}. \quad \text{(A.2)}$$

From the space $H^1(\Omega)$ we can define the fractional order Sobolev space $H^{1/2}(\Gamma)$ as

$$H^{1/2}(\Gamma) = \left\{ f : \Gamma \to \mathbb{R} \mid \exists g \in H^1(\Omega) \text{ so that } g|_\Gamma = f \right\}.$$

The space $H^{-1/2}(\Gamma)$ is the topological dual space of $H^{1/2}(\Gamma)$, that is, the space that contains all the linear and continuous functionals that map the functions of $H^{1/2}(\Gamma)$ to $\mathbb{R}$ (Tartar, 2007).

## Appendix B. The Representation Theorem

The representation theorem allows to represent the solution $u$ of the Laplace's equation on a domain $\Omega$ in terms of its boundary values (Nédélec, 2001).

**Theorem 1.** *Let $\Omega^- \subset \mathbb{R}^3$ be an open, connected set with smooth boundary $\Gamma$, $\Omega^+ = \mathbb{R}^3 \setminus \overline{\Omega^-}$ its complement. Let $\Delta u = 0$ in $\Omega = \Omega^+ \cup \Omega^-$, and let $u$ satisfy the conditions*

$$\lim_{r \to \infty} r |u(\boldsymbol{r})| < \infty, \quad \text{(B.1)}$$

$$\lim_{r \to \infty} r \frac{\partial u}{\partial r}(\boldsymbol{r}) = 0, \quad \text{(B.2)}$$

*where $r = \|\boldsymbol{r}\|$. We define $p = \partial_{\hat{\boldsymbol{n}}} u$. Then it holds*

$$-p = +\mathcal{N}[u] - \mathcal{D}^*[p], \quad \text{for } \boldsymbol{r} \in \Omega, \quad \text{(B.3)}$$

$$u = -\mathcal{D}[u] + \mathcal{S}[p], \quad \text{for } \boldsymbol{r} \in \Omega, \quad \text{(B.4)}$$

$$-p|_\Gamma^\pm = +\mathcal{N}[u] + (\pm \mathcal{I}/2 - \mathcal{D}^*)[p], \quad \text{for } \boldsymbol{r} \in \Gamma, \quad \text{(B.5)}$$

$$u|_\Gamma^\pm = (\mp \mathcal{I}/2 - \mathcal{D})[u] + \mathcal{S}[p], \quad \text{for } \boldsymbol{r} \in \Gamma, \quad \text{(B.6)}$$

*where $\mathcal{I}$ is the identity operator and the operators $\mathcal{S}$, $\mathcal{D}$, $\mathcal{D}^*$, and $\mathcal{N}$ are defined in eqs. (8) to (11).*

Whenever $[u]$ or $[p]$, respectively, are zero, we obtain

$$p|_\Gamma^\pm = \mp [p]/2 + \mathcal{D}^*[p], \quad \text{for } \boldsymbol{r} \in \Gamma, \quad \text{(B.7)}$$

and

$$u|_\Gamma^\pm = \mp [u]/2 - \mathcal{D}[u], \quad \text{for } \boldsymbol{r} \in \Gamma. \quad \text{(B.8)}$$

The representation theorem holds also in the presence of multilayered structures as depicted in fig. 1. Let $\xi_{\Gamma_i}$ and $\mu_{\Gamma_i}$ be functions defined on the $i$th surface. We define the potentials $u_{h1} = \sum_{i=1}^N \mathcal{S} \xi_{\Gamma_i}$ and $u_{h2} = \sum_{i=1}^N \mathcal{D} \mu_{\Gamma_i}$. Then we obtain

$$\partial_{\hat{\boldsymbol{n}}} u_{h1}^\pm(\boldsymbol{r}) = \mp \xi_{\Gamma_j}/2 + \sum_{i=1}^N \mathcal{D}^*_{ji} \xi_{\Gamma_i} \quad \text{for } \boldsymbol{r} \in \Gamma_j, \quad \text{(B.9)}$$

$$u_{h2}^\pm(\boldsymbol{r}) = \pm \mu_{\Gamma_j}/2 + \sum_{i=1}^N \mathcal{D}_{ji} \mu_{\Gamma_i} \quad \text{for } \boldsymbol{r} \in \Gamma_j. \quad \text{(B.10)}$$

## References

Acar, Z. A., Acar, C. E., Makeig, S., 2016. Simultaneous head tissue conductivity and eeg source location estimation. NeuroImage 124, 168–180.

Acar, Z. A., Makeig, S., 2013. Effects of forward model errors on EEG source localization. Brain topography 26 (3), 378–396.

Adde, G., Clerc, M., Faugeras, O., Keriven, R., Kybic, J., Papadopoulo, T., 2003. Symmetric BEM formulation for the m/EEG forward problem. In: Information Processing in Medical Imaging. Springer, pp. 524–535.

Bénar, C.-G., Schön, D., Grimault, S., Nazarian, B., Burle, B., Roth, M., Badier, J.-M., Marquis, P., Liegeois-Chauvel, C., Anton, J.-L., 2007. Single-trial analysis of oddball event-related potentials in simultaneous EEG-fMRI. Human brain mapping 28 (7), 602–613.

Birot, G., Spinelli, L., Vulliémoz, S., Mégevand, P., Brunet, D., Seeck, M., Michel, C. M., 2014. Head model and electrical source imaging: a study of 38 epileptic patients. NeuroImage: Clinical 5, 77–83.

Buffa, A., Christiansen, S., 2007. A dual finite element complex on the barycentric refinement. Mathematics of Computation 76 (260), 1743–1769.

Chan, T. F., 1984. Deflated decomposition of solutions of nearly singular systems. SIAM journal on numerical analysis 21 (4), 738–754.

Clerc, M., Adde, G., Kybic, J., Papadopoulo, T., Badier, J.-M., 2005. In vivo conductivity estimation with symmetric boundary elements. International Journal of Bioelectromagnetism 7, 307–310.

Cools, K., Andriulli, F., De Zutter, D., Michielssen, E., 2011. Accurate and conforming mixed discretization of the mfie. Antennas and Wireless Propagation Letters, IEEE 10, 528–531.

Cools, K., Andriulli, F. P., Olyslager, F., Michielssen, E., 2009. Improving the mfie's accuracy by using a mixed discretization. In: Antennas and Propagation Society International Symposium, 2009. APSURSI'09. IEEE. IEEE, pp. 1–4.

Cosandier-Rimélé, D., Merlet, I., Badier, J.-M., Chauvel, P., Wendling, F., 2008. The neuronal sources of EEG: modeling of simultaneous scalp and intracerebral recordings in epilepsy. NeuroImage 42 (1), 135–146.

Dabek, J., Kalogianni, K., Rotgans, E., van der Helm, F. C., Kwakkel, G., van Wegen, E. E., Daffertshofer, A., de Munck, J. C., 2015. Determination of head conductivity frequency response in vivo with optimized eit-eeg. NeuroImage.

De Munck, J., 1988. The potential distribution in a layered anisotropic spheroidal volume conductor. Journal of applied Physics 64 (2), 464–470.

De Munck, J., 1992. A linear discretization of the volume conductor boundary integral equation using analytically integrated elements. IEEE transactions on bio-medical engineering 39 (9), 986–990.

De Munck, J., Van Dijk, B., Spekreijse, H., 1988. Mathematical dipoles are adequate to describe realistic generators of human brain activity. Biomedical Engineering, IEEE Transactions on 35 (11), 960–966.

Despotovic, I., Vansteenkiste, E., Philips, W., 2013. A realistic volume conductor model of the neonatal head: Methods, challenges and applications. In: Engineering in Medicine and Biology Society (EMBC), 2013 35th Annual International Conference of the IEEE. IEEE, pp. 3303–3306.

Dunavant, D., 1985. High degree efficient symmetrical gaussian quadrature rules for the triangle. International journal for numerical methods in engineering 21 (6), 1129–1148.

Ernst, L. M., Ruchelli, E. D., Huff, D. S., 2011. Color atlas of fetal and neonatal histology. Springer.

Fedorov, A., Beichel, R., Kalpathy-Cramer, J., Finet, J., Fillion-Robin, J.-C., Pujol, S., Bauer, C., Jennings, D., Fennessy, F., Sonka, M., et al., 2012. 3d slicer as an image computing platform for the quantitative imaging network. Magnetic resonance imaging 30 (9), 1323–1341.

Fiederer, L., Vorwerk, J., Lucka, F., Dannhauer, M., Yang, S., Dümpelmann, M., Schulze-Bonhage, A., Aertsen, A., Speck, O., Wolters, C., et al., 2016. The role of blood vessels in high-resolution volume conductor head modeling of eeg. NeuroImage 128, 193–208.

Fischl, B., van der Kouwe, A., Destrieux, C., Halgren, E., Ségonne, F., Salat, D. H., Busa, E., Seidman, L. J., Goldstein, J., Kennedy, D., et al., 2004. Automatically parcellating the human cerebral cortex. Cerebral cortex 14 (1), 11–22.

Fuchs, M., Drenckhahn, R., Wischmann, H., Wagner, M., 1998. An improved boundary element method for realistic volume-conductor modeling. Biomedical Engineering, IEEE Transactions on 45 (8), 980–997.

Fuchs, M., Wagner, M., Kastner, J., 2001. Boundary element method volume conductor models for EEG source reconstruction. Clinical neurophysiology 112 (8), 1400–1407.

Geffroy, D., Rivière, D., Denghien, I., Souedet, N., Laguitton, S., Cointepas, Y., 2011. Brainvisa: a complete software platform for neuroimaging. In: Python in Neuroscience Workshop. Euroscipy Paris.

Gençer, N. G., Akalin-Acar, Z., 2005. Use of the isolated problem approach for multi-compartment bem models of electro-magnetic source imaging. Physics in medicine and biology 50 (13), 3007.

Goebel, R., 2012. Brainvoyager—past, present, future. Neuroimage 62 (2), 748–756.

Gonçalves, S., De Munck, J. C., Verbunt, J., Bijma, F., Heethaar, R. M., Lopes da Silva, F., et al., 2003. In vivo measurement of the brain and skull resistivities using an eit-based method and realistic models for the head. Biomedical Engineering, IEEE Transactions on 50 (6), 754–767.

Graglia, R. D., 1987. Static and dynamic potential integrals for linearly varying source distributions in two-and three-dimensional problems. Antennas and Propagation, IEEE Transactions on 35 (6), 662–669.

Graglia, R. D., 1993. On the numerical integration of the linear shape functions times the 3-d green's function or its gradient on a plane triangle. Antennas and Propagation, IEEE Transactions on 41 (10), 1448–1455.

Gramfort, A., Luessi, M., Larson, E., Engemann, D. A., Strohmeier, D., Brodbeck, C., Parkkonen, L., Hämäläinen, M. S., 2014. Mne software for processing meg and eeg data. Neuroimage 86, 446–460.

Grech, R., Cassar, T., Muscat, J., Camilleri, K. P., Fabri, S. G., Zervakis, M., Xanthopoulos, P., Sakkalis, V., Vanrumste, B., 2008. Review on solving the inverse problem in EEG source analysis. Journal of neuroengineering and rehabilitation 5 (1), 25.

Hackbusch, W., 2012. Integral equations: theory and numerical treatment. Vol. 120. Birkhäuser.

Hallez, H., Vanrumste, B., Grech, R., Muscat, J., De Clercq, W., Vergult, A., D'Asseler, Y., Camilleri, K. P., Fabri, S. G., Van Huffel, S., et al., 2007a. Review on solving the forward problem in EEG source analysis. Journal of neuroengineering and rehabilitation 4 (1), 46.

Hallez, H., Vanrumste, B., Grech, R., Muscat, J., De Clercq, W., Vergult, A., D'Asseler, Y., Camilleri, K. P., Fabri, S. G., Van Huffel, S., et al., 2007b. Review on solving the forward problem in EEG source analysis. Journal of neuroengineering and rehabilitation 4 (1), 46.

Hamalainen, M. S., Sarvas, J., 1989. Realistic conductivity geometry model of the human head for interpretation of neuromagnetic data. Biomedical Engineering,

IEEE Transactions on 36 (2), 165–171.

He, B., Musha, T., Okamoto, Y., Homma, S., Nakajima, Y., Sato, T., 1987. Electric dipole tracing in the brain by means of the boundary element method and its accuracy. Biomedical Engineering, IEEE Transactions on (6), 406–414.

He, B., Wang, Y., Wu, D., 1999. Estimating cortical potentials from scalp EEGs in a realistically shaped inhomogeneous head model by means of the boundary element method. Biomedical Engineering, IEEE Transactions on 46 (10), 1264–1268.

Huang, Y., Parra, L. C., Haufe, S., 2015. The new york head—a precise standardized volume conductor model for eeg source localization and tes targeting. NeuroImage.

Huiskamp, G., Vroeijenstijn, M., van Dijk, R., Wieneke, G., van Huffelen, A. C., 1999. The need for correct realistic geometry in the inverse EEG problem. Biomedical Engineering, IEEE Transactions on 46 (11), 1281–1287.

Jorge, J., Grouiller, F., Gruetter, R., Van Der Zwaag, W., Figueiredo, P., 2015. Towards high-quality simultaneous eeg-fmri at 7t: Detection and reduction of eeg artifacts due to head motion. NeuroImage 120, 143–153.

Koessler, L., Benar, C., Maillard, L., Badier, J.-M., Vignal, J. P., Bartolomei, F., Chauvel, P., Gavaret, M., 2010. Source localization of ictal epileptic activity investigated by high resolution EEG and validated by sEEG. Neuroimage 51 (2), 642–653.

Kybic, J., Clerc, M., Abboud, T., Faugeras, O., Keriven, R., Papadopoulo, T., 2005. A common formalism for the integral formulations of the forward EEG problem. Medical Imaging, IEEE Transactions on 24 (1), 12–28.

Meijs, J. W., Weier, O. W., Peters, M. J., van Oosterom, A., 1989. On the numerical accuracy of the boundary element method (EEG application). Biomedical Engineering, IEEE Transactions on 36 (10), 1038–1049.

Munck, d. J., Peters, M. J., 1993. A fast method to compute the potential in the multisphere model. IEEE transactions on biomedical engineering 40 (11), 1166–1174.

Nédélec, J.-C., 2001. Acoustic and electromagnetic equations: Integral representations for harmonic problems. Vol. 144. Springer.

Odabaee, M., Layeghy, S., Mesbah, M., Azemi, G., Boashash, B., Vanhatalo, S., et al., 2012. EEG amplitude and correlation spatial decay analysis for neonatal head modelling. In: Information Science, Signal Processing and their Applications (ISSPA), 2012 11th International Conference on. IEEE, pp. 882–887.

Odabaee, M., Tokariev, A., Layeghy, S., Mesbah, M., Colditz, P. B., Ramon, C., Vanhatalo, S., 2014. Neonatal EEG at scalp is focal and implies high skull conductivity in realistic neonatal head models. NeuroImage 96, 73–80.

Oostendorp, T. F., Delbeke, J., Stegeman, D. F., 2000. The conductivity of the human skull: results of in vivo and in vitro measurements. Biomedical Engineering, IEEE Transactions on 47 (11), 1487–1492.

Oostenveld, R., Fries, P., Maris, E., Schoffelen, J.-M., 2010. Fieldtrip: open source software for advanced analysis of MEG, EEG, and invasive electrophysiological data. Computational intelligence and neuroscience 2011.

Pascual-Marqui, R. D., 1999. Review of methods for solving the EEG inverse problem. International journal of bioelectromagnetism 1 (1), 75–86.

Patrizi, S., Holmes, G. L., Orzalesi, M., Allemand, F., 2003. Neonatal seizures: characteristics of EEG ictal activity in preterm and fullterm infants. Brain and Development 25 (6), 427–437.

Peng, K., Nguyen, D. K., Vannasing, P., Tremblay, J., Lesage, F., Pouliot, P., 2016. Using patient-specific hemodynamic response function in epileptic spike analysis of human epilepsy: a study based on eeg–fnirs. NeuroImage 126, 239–255.

Phillips, C., Rugg, M. D., Friston, K. J., 2002. Systematic regularization of linear inverse solutions of the EEG source localization problem. NeuroImage 17 (1), 287–301.

Pruis, G., Gilding, B. H., Peters, M., 1993. A comparison of different numerical methods for solving the forward problem in EEG and MEG. Physiological measurement 14 (4A), A1.

Rahmouni, L., Andriulli, F., April 2014. Mixed discretization formulations for the direct EEG problem. In: Antennas and Propagation (EuCAP), 2014 8th European Conference on. pp. 3183–3185.

Rahola, J., Tissari, S., 2002. Iterative solution of dense linear systems arising from the electrostatic integral equation in MEG. Physics in medicine and biology 47 (6), 961.

Sarvas, J., 1987. Basic mathematical and electromagnetic concepts of the biomagnetic inverse problem. Physics in medicine and biology 32 (1), 11.

Sauter, S. A., Schwab, C., 2011. Boundary element methods. Springer.

Scherzer, O., 2011. Handbook of Mathematical Methods in Imaging: Vol. 1. Springer Science & Business Media.

Schimpf, P. H., Ramon, C., Haueisen, J., 2002. Dipole models for the EEG and MEG. Biomedical Engineering, IEEE Transactions on 49 (5), 409–418.

Shattuck, D. W., Leahy, R. M., 2002. Brainsuite: an automated cortical surface identification tool. Medical image analysis 6 (2), 129–142.

Siems, M., Pape, A.-A., Hipp, J. F., Siegel, M., 2016. Measuring the cortical correlation structure of spontaneous oscillatory activity with eeg and meg. NeuroImage.

Silau, A. M., Fischer, H. B., Kjær, I., 1994. Normal prenatal development of the human parietal bone and interparietal suture. Journal of craniofacial genetics and developmental biology 15 (2), 81–86.

Smith, S. M., 2002. Fast robust automated brain extraction. Human brain mapping 17 (3), 143–155.

Steinbach, O., 2008. Numerical approximation methods for elliptic boundary value problems. Finite and Boundary Elements.

Steinbach, O., Wendland, W. L., 1998. The construction of some efficient preconditioners in the boundary element method. Advances in Computational Mathematics 9 (1-2), 191–216.

Stenroos, M., Sarvas, J., 2012. Bioelectromagnetic forward problem: isolated source approach revis (it) ed. Physics in medicine and biology 57 (11), 3517.

Tartar, L., 2007. An Introduction to Sobolev Spaces and Interpolation. Springer.

Watanabe, K., Hayakawa, F., Okumura, A., 1999. Neonatal EEG: a powerful tool in the assessment of brain damage in preterm infants. Brain and Development 21 (6), 361–372.

Wolters, C. H., Köstler, H., Möller, C., Härdtlein, J., Grasedyck, L., Hackbusch, W., 2007. Numerical mathematics of the subtraction method for the modeling of a current dipole in EEG source reconstruction using finite element head models. SIAM Journal on Scientific Computing 30 (1), 24–45.

Yan, S., Jin, J.-M., Nie, Z., 2011. Improving the accuracy of the second-kind fredholm integral equations by using the a-christiansen functions. Antennas and Propagation, IEEE Transactions on 59 (4), 1299–1310.

Ylä-Oijala, P., Kiminki, S. P., Järvenpää, S., 2015. Conforming boundary element methods in acoustics. Engineering Analysis with Boundary Elements 50, 447–458.

Yvert, B., Crouzeix-Cheylus, A., Pernier, J., 2001. Fast realistic modeling in bio-electromagnetism using lead-field interpolation. Human brain mapping 14 (1), 48–63.

Zanow, F., Peters, M., 1995. Individually shaped volume conductor models of the head in EEG source localisation. Medical and biological engineering and computing 33 (4), 582–588.

Zhang, Y., van Drongelen, W., He, B., 2006. Estimation of in vivo brain-to-skull conductivity ratio in humans. Applied physics letters 89 (22), 223903.

Zhang, Z., 1995. A fast method to compute surface potentials generated by dipoles within multilayer anisotropic spheres. Physics in medicine and biology 40 (3), 335.